\documentclass[11pt,twoside]{article}
\usepackage{./asp2010}
\resetcounters
\begin{document}

\title{Scaling laws in disk galaxies}
\author{Alister W.\ Graham$^1$
\affil{$^1$Centre for Astrophysics and Supercomputing, Swinburne University of Technology, Hawthorn, Victoria 3122, Australia.}}

\begin{abstract}
  A brief overview of several recent disk galaxy scaling relations is
  presented, along with some historical background.  In particular, after
  introducing the (basic) radial light profiles of disk galaxies, I explain 
  several important `structural' scaling relations and review the latest bulge-(black
  hole mass) diagrams.  I go on to present the typical bulge-to-disk flux 
  ratios of disk galaxies and suggest the use of a grid of bulge-to-disk ratio
  versus disk Hubble type --- based on the nature of the spiral arms --- to
  complement the Hubble-Jeans sequence.  I then briefly mention pure disk
  galaxies before cautioning on difficulties with identifying pseudobulges
  built from secular evolution.  Finally, I conclude by discussing a likely
  connection between modern day bulges in disk galaxies and high-redshift
  ($z\sim 2 \pm 0.5$) compact galaxies which may have since acquired a disk
  via cold flows and quiescent accretion.
\end{abstract}


\section{Light profiles --- the radial concentration of stars}

Bulges are centrally-located stellar distributions with a smooth appearance
that manifest themselves as an excess, a bulge, relative to the inward
extrapolation of their galaxies outer disk light (Wyse et al.\ 1997); noting a
distinction between obvious bars, nuclear disks and star clusters.  The bulk
of a disk can be well approximated by the exponential model (Patterson 1940;
de Vaucouleurs 1957, 1958; Freeman 1970).  In general, they may be comprised
of a thick, thin, and super-thin disk (Schechtman-Rook 2013), often display a
break and/or truncation at large radii (van der Kruit 1979, 1987; Erwin,
Pohlen \& Beckman 2008; Hermann 2013), and bars of varying strength appear in
some two-thirds of the local disk galaxy population (e.g.\ Sheth et al.\
2013).

Bulges were, for a long time, considered to have $R^{1/4}$ light profiles
(e.g.\ de Vaucouleurs 1958; Kormendy 1977a; Burstein 1979; Kent 1985; Kodaira
et al.\ 1986; Simien \& de Vaucouleurs 1986; Baggett et al.\ 1998; Lilly et
al.\ 1998; etc.). So popular was the belief in this model that it was referred
to as the $R^{1/4}$ {\it law}.  However, de Vaucouleurs (1959) had noted
departures in some bulge light profiles from his (1948) $R^{1/4}$ model, and
the exponential model was subsequently shown by others to provide better fits
for some bulges (e.g.\ van Houten 1961; Liller 1966; Frankston \& Schild 1976;
Spinrad et al.\ 1978).  Shaw \& Gilmore (1989) and Wainscoat et al.\ (1989)
reiterated that not all bulges are well described with the $R^{1/4}$ model,
and Andredakis \& Sanders (1994) showed that {\it many} bulges are in fact
better fit with an exponential model.  Andredakis, Peletier \& Balcells (1995)
subsequently revealed the suitability of S\'ersic's (1963, 1968) $R^{1/n}$
model, which could accommodate $R^{1/4}$ profiles, exponential ($n=1$)
profiles, plus everything in between and at either extreme (i.e.\ $n<1$ and
$n>4$).  Furthermore, building on the work with elliptical galaxies by Caon et
al.\ (1993) and Young \& Currie (1994), Andredakis et al.\ provided valuable
insight by showing that the bulge S\'ersic index correlates with the absolute
magnitude of the bulge.  Moreover, Balcells et al.\ (2003) revealed that most
early-type disk galaxies do not actually have $R^{1/4}$ bulges but instead
light profiles which have S\'ersic indices $n < 4$.

This year is the golden anniversary of S\'ersic's model, which 
was presented in Spanish by Jose S\'ersic 50 years ago.  His highly useful
$R^{1/n}$ model is reviewed in Graham \& Driver (2005).  Briefly, 
the radial surface brightness profiles (see Fig.~\ref{Fig1}) are such that 
\begin{equation}
\mu (R)=\mu_{\rm e}+\frac{2.5b_n}{\ln(10)}\left[\left(R/R_{\rm
      e}\right)^{1/n}-1\right], 
\end{equation}
where $\mu_{\rm e}$ is the effective surface brightness at the effective half
light radius $R_{\rm e}$.  The term 
$b_n$ is not a parameter but a function of the S\'ersic index $n$ such that $b_n \approx
2n -1/3$, which ensures that $R_{\rm e}$ encloses half of the model's total light. 
The average, or mean, effective surface brightness within $R_{\rm e}$ is such
that 
$\langle\mu\rangle_{\rm e} = \mu_{\rm e} -2.5\log[n {\rm e}^{b_n}\Gamma (2n) /
(b_n)^{2n}]$, 
and the total luminosity $L=2\pi R_{\rm e}^2\langle I \rangle_{\rm e}$, 
where $\langle\mu\rangle_{\rm e} = -2.5\log\langle I \rangle_{\rm e}$. 
Importantly, the effective surface brightness is related to the central
surface brightness $\mu_0$ in a non-linear fashion such that 
\begin{equation}
\mu_0 = \mu_{\rm e} -2.5b_n/\ln(10). \label{Eq_mu}
\end{equation}

The impact that the range of bulge light profile shapes has, such as how
$\mu_{\rm e} - \mu_0$ and $\langle\mu\rangle_{\rm e} - \mu_0$ vary with $n$,
is yet to be fully appreciated within the community.  It turns out that this
broken homology, i.e.\ the systematic departures from the $R^{1/4}$ model as a
function of bulge magnitude, has a dramatic 
influence on the scaling relations we construct using the effective
parameters.

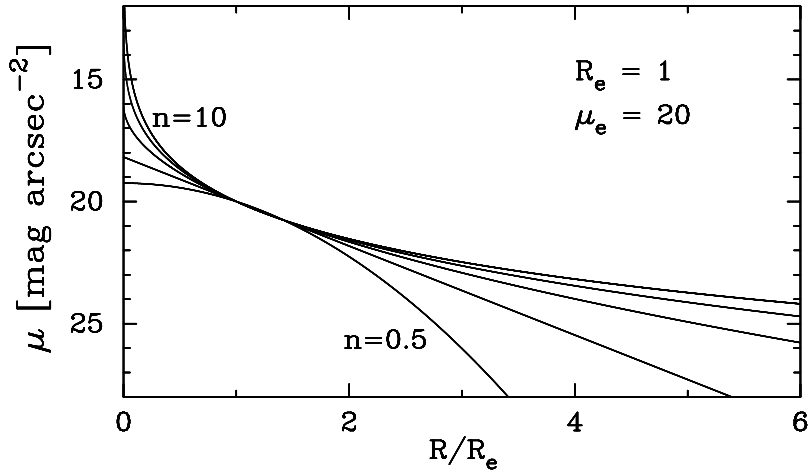
\articlefigure[angle=-90,scale=0.75]{grahama-fig1.ps}{Fig1}{S\'ersic $R^{1/n}$
  surface brightness profiles.  Figure adapted from Graham \& Driver (2005).}

\section{A few select disk galaxy scaling relations}

\subsection{Insight from elliptical galaxies}

Although the bulges of disk galaxies are not simply small elliptical galaxies
--- because their stellar densities are much higher (see Section~\ref{Sec6})
--- they are similar in the sense that the 3-parameter S\'ersic model can
describe their light profiles.  One can use the luminosity (or rather the
absolute magnitude $M = -2.5\log[L]$), central surface brightness and S\'ersic
index as the three independent variables.  From a sample of many elliptical
galaxies, Fig.~\ref{Fig2} reveals that these parameters follow log-linear
relations that unify the dwarf and giant elliptical galaxies across the
alleged divide at $M_B = -18$ mag (Graham \& Guzm\'an 2003).  However, when
one uses the effective half light parameters (Fig.~\ref{Fig3}), the
continuous, unifying relations are curved. The effective half light radius,
and associated surface brightness, have always been somewhat arbitrary
quantities.  For example, we (astronomers) could have used a radius encolsing
25\% or 75\% of the total light, and these curved relations would look
different again. This would not happen if every light profile was actually an
$R^{1/4}$ profile.  Without this understanding, it had led some to conclude
that the dwarf and giant elliptical galaxies must be distinct species due to
their apparent, near-orthogonal distribution in diagrams involving the
effective parameters.  Fig.~\ref{Fig3} reveals, for example, how the
log-linear $M-\mu_0$ relation (Fig.~\ref{Fig2}a) maps into the curved
$M-\mu_{\rm e}$ relation due to the difference between $\mu_0$ and $\mu_{\rm
  e}$ seen in Fig.~\ref{Fig1} and described by Eq.~\ref{Eq_mu}.  It also
reveals the highly curved nature of the $\langle\mu\rangle_{\rm e}$--$R_{\rm
  e}$ distribution.  As detailed in Graham (2013), this curve shown in
Fig.~\ref{Fig3}c has been derived from the log-linear relations that unify
the faint and bright elliptical galaxies in Fig.~\ref{Fig2}.

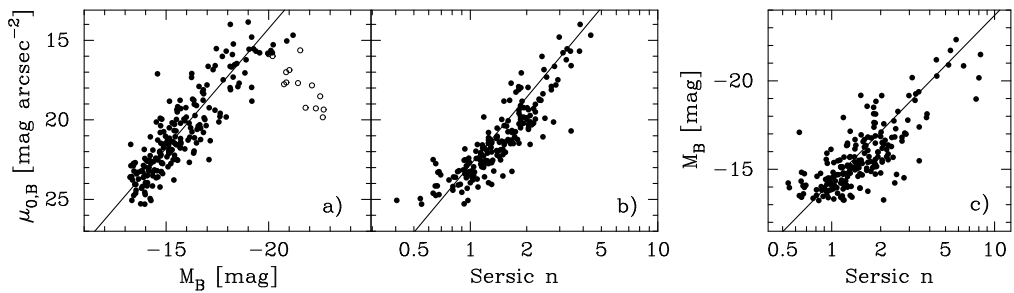
\articlefigure[angle=-90,scale=0.55]{grahama-fig2.ps}{Fig2}{The ($B$-band)
  magnitude ($M = -2.5\log[L]$), central surface brightness ($\mu_0$) and
  S\'ersic index ($n$) of elliptical galaxies define unifying, continuous,
  log-linear relations.  The deviation at the bright-end of the $\mu_0$--$M$
  relation in panel a) is due to the presence of partially depleted cores,
  thought to have formed from the scouring of binary black holes (e.g.\
  Begelman et al.\ 1980; Faber et al.\ 1997; Dullo \& Graham 2013, and
  references therein).}

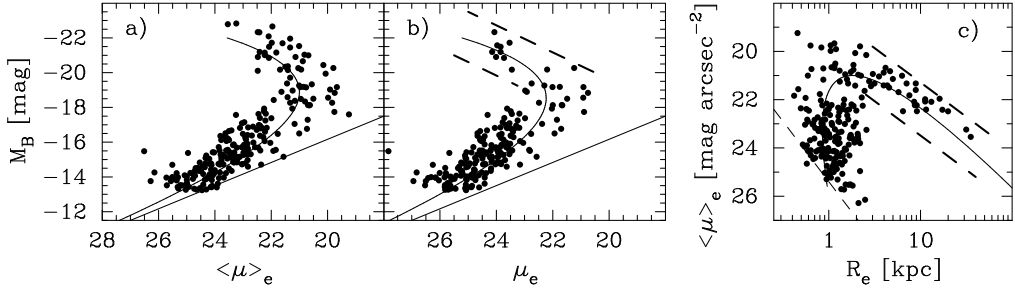
\articlefigure[angle=-90,scale=0.54]{grahama-fig3.ps}{Fig3}{The linear
  relations from Fig.~\ref{Fig2} map into curved relations when `effective'
  parameters (such as $R_{\rm e}, \mu_{\rm e}$ or $\langle\mu\rangle_{\rm e}$)
  are involved. The bright-arms of these curved relations, marked with the
  dashed lines, are known as the Kormendy (1977b) relation.  Figure adapted from 
  Graham \& Guzm\'an (2003) and Graham
  (2013).}

\subsection{Bulges}\label{Sec22}

We can use the above insight from elliptical galaxies to better understand the
bulges of disk galaxies.  Just as there is no dichotomy between dwarf
elliptical and giant elliptical galaxies at $M_B = -18$ mag, or roughly
S\'ersic $n=2$, 
Fig.~\ref{Fig4} reveals no sign of a dichotomy between `classical' $n=4$ bulges
and $n=1$ (or $n\le2$) `pseudobulges'. 
One can also see that, as with the elliptical galaxies, the continuous and linear 
$M-n$ and $M-\mu_0$ relations for bulges 
map into a continuous but curved $M-\mu_{\rm e}$
relation (Fig.~\ref{Fig4}c). 

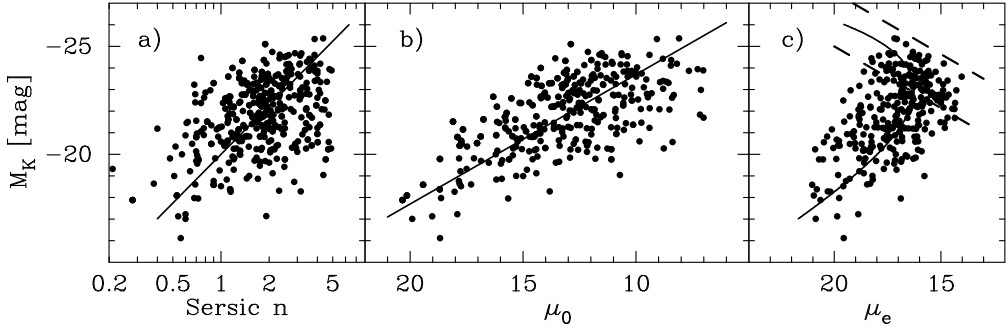
\articlefigure[angle=-90,scale=0.58]{grahama-fig4.ps}{Fig4}{Similar to
  elliptical galaxies, bulges follow single log-linear scaling relations involving
  absolute magnitude, central surface brightness and S\'ersic index. 
  These two log-linear relations shown in panels a) and b) map into the 
  curved relation shown in panel
  c), which is not a fit to the data shown there but the requirement /
  prediction from these two log-linear relations.  $K$-band
  data has been used here from the compilation in Graham \& Worley (2008). 
  The bright arm of the curved (absolute 
  magnitude)-(effective surface brightness) relation for bulges in panel c) 
  is delineated by the dashed lines.  Figure adapted from Graham (2013).} 

Although bulges with $n \lesssim 2$ will `appear' to deviate from those with $n \gtrsim 2$ in the
$M$–-$\mu_{\rm e}$ diagram, the $R_{\rm e}$-–$\mu_{\rm e}$ diagram, and also the 
Fundamental Plane (as noted in Graham \& Guzm\'an 2004, see also Guzm\'an,
Graham \& Ismay, in prep.), this is not evidence of a dichotomy but
simply reflects that the distribution is curved in diagrams involving
effective parameters 
(see also Graham \& Guzm\'an 2003; Gavazzi et al.\ 2005; Ferrarese et al.\
2006; C\^ot\'e et al.\ 2006, 2007). 

While rotating bulges with $n\sim1$ may be a sign of `pseudobulges' grown from
the secular evolution of a disk, some care 
is needed because Dom{\'{\i}}nguez-Tenreiro et al.\ (1998), Aguerri et al.\ (2001) and
Scannapieco et al.\ (2010) have grown bulges with $1 < n < 2$ from minor mergers.
Further evidence that a radial light distribution with an exponential profile
need not have formed from the secular evolution of a disk are the existence of
both dwarf elliptical galaxies (Young \& Currie 1994) and cD galaxy halos
(Seigar et al.\ 2007; Pierini et al.\ 2008) which have light profiles with $n\approx1$.

\section{Bulge-to-disk flux ratios and the Hubble sequence}

The Hubble-Jeans sequence (Jeans 1919, 1928; Hubble 1926, 1936; 
reviewed in van den Bergh 1997 and Sandage 2004) is not 
purely a bulge-to-disk ($B/D$) flux ratio sequence; the pitch angle 
(e.g., Kennefick 2013; Puerari 2013; Davis 2013) is a primary criteria
for determining the Hubble type.  Indeed, in the Hubble Atlas of Galaxies,
Sandage (1961) made the spiral arms the primary criteria, noting that Sa galaxies can
therefore exist with both large and small bulges.
There is in fact a range of $B/D$ flux ratios for any given spiral 
Hubble type (Fig.~\ref{Fig5}). 

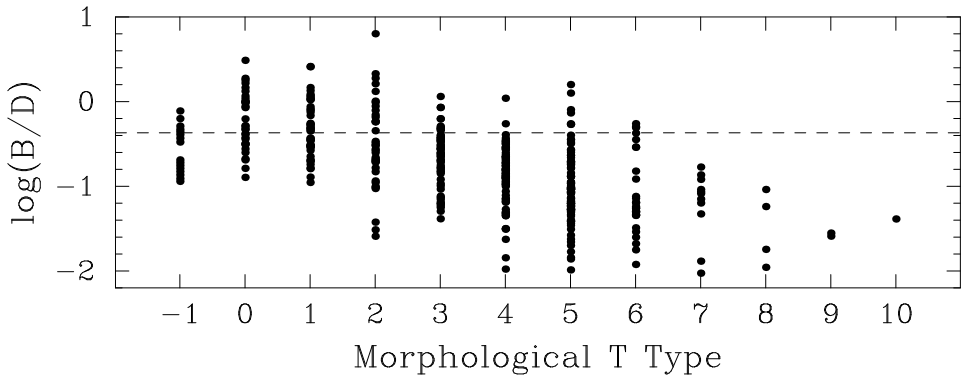
\articlefigure[angle=-90,scale=0.52]{grahama-fig5.ps}{Fig5}{Bulge-to-disk
  flux ratios. Adapted from Graham \& Worley (2008). The dashed line
  corresponds to a bulge-to-total flux ratio of 0.3, and reveals  no evidence
for a divide which has been advocated by some as delineating pseudobulges from
classical bulges. (Note: T=1=Sa, T=2=Sab, T=3=Sb, T=5=Sc, T=7=Sd, etc.)}

The range of $B/D$ flux ratios for the lenticular galaxies 
had led some to propose an S0 sequence of varying $B/D$ flux ratio, running 
parallel to the spiral sequence (van den Bergh 1976; Cappellari et al.\ 2011;
Kormendy \& Bender 2012), but as Fig.~\ref{Fig5} shows, the 
spiral galaxies actually occupy a grid of pitch angle (i.e.\ roughly galaxy type: Sa,
Sb, Sc, Sd) verus $B/D$ ratio. 
That is, it not simply that the S0 galaxies display a range or sequence
of $B/D$ values, every disk galaxy type does. 
It is therefore suggested here 
that a grid (of spiral type based on the spiral arms, 
and the bulge-to-disk flux ratio or bulge magnitude) 
might be a useful ($z=0$) classification scheme, in addition to the Hubble-Jeans 
sequence and expanding upon the van den Bergh track of just three parallel
sequences, referred to as a comb diagram by Cappellari et al.\ (2011). 
That is, while this comb diagram focusses on the bulge-to-disk flux ratio and
the prominance of the spiral arms (strong for spirals, non-existent for
lenticular galaxies, and of intermediate strength for the anemic spirals),
there may also be value in a grid which focusses on the bulge-to-disk flux ratio 
and the extent to which the spiral arms are unwound.

\subsection{Bulgeless galaxies}

There is some need for qualification when it comes to the term `bulgeless'. 
The Milky Way has been referred to as a bulgeless galaxy by some who regard it as a 
pure disk galaxy harboring no `classical' bulge. 
However, it certainly has a bulge relative to the underlying disk, 
and D\'ek\'any et al.\ (2013) 
suggest that it is comprised of both a classical bulge and a pseudobulge. 
The Scd galaxy NGC~1042 is another example which has recently been heralded as a bulgeless
galaxy, and importantly one with an AGN (i.e.\ a supermassive black hole), but Knapen et 
al.\ (2003) have shown that it does actually have a bulge. 
Similarly, Simmons et al.\ (2013) call galaxies bulgeless if $B/T$ is small ($<
\sim$0.05), even though a bulge is often still apparent. 
There is thus some ambiguity as to what is meant by `bulgeless'. 
Having noted this, readers are directed to the references in Secrest et al.\
(2013) for 'bulgeless' galaxies with AGN, some of which may truly be
bulgeless.

While IC~5249 has less than 2\% of its total light in a bulge component, 
NGC~300 (Bland-Hawthorn et al.\ 2005) has no perceivable bulge. 
There are also examples of apparently bulgeless, super-thin edge-on 
galaxies (Kautsch 2009), 
and super-thin disks such as in NGC~891 (see Schechtman-Rook 2013).

\section{(Black hole)-bulge scaling relations}

Within spiral galaxies, one may ask if the mass of the central 
black hole scales better with the bulge mass (or bulge-to-disk ratio), or 
perhaps some other property such as the pitch angle of the spiral arms
(e.g.\ Seigar et al.\ 2008; Treuthardt et al.\ 2012; Berrier et al.\ 2013). 
If truly bulgeless galaxies do have active galactic nuclei (AGN), then the $M_{\rm
  bh}$-$M_{\rm bulge}$ mass relation defined by big bulges can not apply for 
obvious reasons.  The existence of spiral patterns in pure disk galaxies
(e.g.\ Valencia-Enriquez \& Puerari 2013) at least offers hope for an $M_{\rm
  bh}$-(pitch angle) relation in disks with AGN but no bulge. 
However the possibly transient nature of spiral arms raises the question as 
to the time scale on which spiral arms may evolve (Minchev et al.\ 2012;
Martinez-Garcia 2013; Shields 2013), 
and does this evolution trace the growth of the black holes (and bulges)? 
It may be of value to explore if the amplitude/contrast of the spiral arms
(Grosb{\o}l 2013) is important for the above mentioned $M_{\rm bh}$-(pitch
angle) relation given the existence of a) dwarf galaxies with very faint spiral
arms (e.g.\ Jerjen et al.\ 2000; Graham et al.\ 2003), and b) the anemic spirals
in the classification scheme of van den Bergh (1976).

In Fig.~\ref{Fig6}a, the bulges of barred galaxies appear to house black holes
which are, on average, a factor of two less massive than those in non-barred
galaxies of the same velocity dispersion (Graham 2008a,b, Hu 2008; Graham et
al.\ 2011).  However, as was initially speculated, this result is instead 
due to the bar's dynamics having elevated the velocity dispersion (Hartmann et
al.\ 2013, see also Brown et al.\ 2013). That is, the barred galaxies are 
offset in velocity dispersion rather than black hole mass.

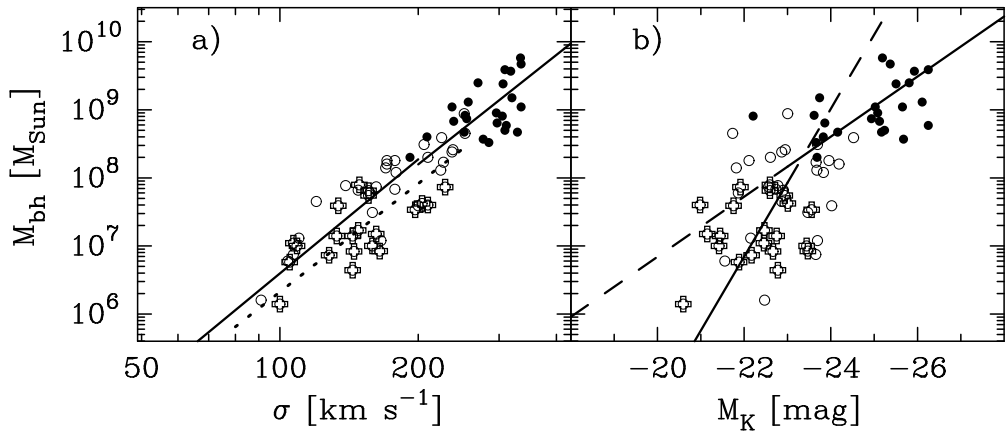
\articlefigure[angle=-90,scale=0.8]{grahama-fig6.ps}{Fig6}{Black hole mass
  versus a) velocity dispersion and b) $K$-band bulge magnitude.  Galaxies
  with depleted cores are shown by the filled circles. Galaxies without
  depleted cores are shown by the open symbols.  Barred galaxies are shown by
  the crosses.  Separate linear regressions have been applied to all three
  types in panel a), while the latter two types have been grouped together in
  panel b) due to the level of scatter in the current data (see Graham \&
  Scott 2013 for more details).}

It is not yet established if barred galaxies follow a different distribution
in the $M_{\rm bh}$-$L_{\rm bulge}$ diagram (Fig.~\ref{Fig6}b).  
One important, recent realisation is that because bulges of (barred
and non-barred) disk galaxies follow the $M_{\rm bh} \propto
\sigma^5$ scaling relation (Fig.~\ref{Fig6}a), and the luminosity $L \propto \sigma^2$ for 
bulges with absolute B-band magnitudes fainter than $M_B = -20.5\pm1$ mag (e.g.\ Davies et al.\ 1983; 
Matkovi\'c \& Guzm\'an 2005; Kourkchi et al.\ 2012), it requires that 
$M_{\rm bh} \propto L^{2.5}$ (or $M_{\rm bh} \propto M^{2}_{\rm bulge,dyn}$, if
$M_{\rm dyn}/L \propto L^{1/4}$)\footnote{Note: Cappellari et al.\ (2006) report
  $M_{\rm dyn}/L \propto L^{1/3}$, and the $M_{\rm dyn}$--$\sigma$ relation from 
Cappellari et al.\ (2013, their figure~1) appears to show a bend at roughly 
$10^{11} M_{\odot}$, in the middle of the distribution where $\log[M_{\rm
  dyn}/L]=0.75$.}.  That is, there should not be a linear $M_{\rm bh}$-$M_{\rm
  bulge}$ relation for these galaxies, but rather a 
near-quadratic relation, and this is what is observed in Fig.~\ref{Fig6}b 
(Graham 2012; Graham \& Scott 2013; Scott et al.\ 2013). 


\section{Pseudobulges}

`Pseudobulges' are supposed to rotate and have an exponential light profile,
akin to the disk material from which they formed (Bardeen 1975; Hohl 1975;
Hohl \& Zhang 1979; Combes \& Sanders 1981).  The topic of their light
profiles was already addressed in Section~\ref{Sec22} with a large galaxy
sample, where it was both explained why and shown that there is no physical
divide at $n=2$ in the structural diagrams, although see Fisher \& Drory
(2008) for an alternative view.  
It was also pointed out that the scaling relations involving the
`effective' structural parameters are curved and therefore cannot on their own
be used to identify different bulge (formation) type.  


\subsection{Rotation}

Bulges have of course been known to rotate for many years 
(e.g. Pease 1918; Babcock 1938; Rubin, Ford \& Kumar 1973).  

Merger events can create rotating elliptical galaxies 
(e.g.\ Naab, Khochfar \& Burkert 2006; 
Gonz\'alez-Garc{\'{\i}}a et al.\ 2009; Hoffman et al.\ 2009), and 
merger simulations can also create bulges which rotate (Bekki 2010;
Keselman \& Nusser 2012). 
Classical bulges can be spun up by a bar (Saha et al.\ 2012).
Bar dynamics may give the illusion of rotation in classical bulges
(Babusiaux et al.\ 2010). 
Williams et al.\ (2010) report that some boxy bulges, (previously) thought to be
bars seen in projection (Combes \& Sanders 1981), do not 
display cylindrical rotation as expected and can have stellar populations
different to their disk.
Qu et al.\ (2011) report on how the rotational delay between old
and young stars in the disk of our Galaxy may be a signature of a
minor merger event.

For the above reasons, rotation is not a definitive sign of bulges built via
secular disk processes, and as such it can not be used to definitively
identify bulge type.

\subsection{Ages}

From optical and near-IR colours, 
Peletier et al.\ (1999) concluded (after avoiding dusty regions)
that the bulges of S0-Sb galaxies are old and cannot have
formed from secular evolution more recently than z = 3.
Bothun \& Gregg (1990) had previously argued that bulges in
S0 galaxies are typically 5 Gyr older than their disks.
Bell \& de Jong (2000) reported that bulges tend to be older
and more metal rich than disks in all galaxy types, and
Carollo et al.\ (2007) found that roughly half of their late-type
spirals had old bulges.
Gadotti \& dos Anjos (2001) found that $\approx$60\% of Sbc galaxies
have bulge colours which are redder than their disks. 
For reference, it is noted that the average Sbc spiral has a S\'ersic index $n
< 2$ (Graham \& Worley 2008). 

From spectra, Goudfrooij, Gorgas \& Jablonka (1999) reported that bulges in
their sample of edge-on spiral galaxies are old (like in elliptical galaxies), and
have super-solar $\alpha$/Fe ratios similar to those of giant elliptical galaxies.  They
concluded that their observations favor the `dissipative collapse'
model rather than the `secular evolution' model. 
Thomas \& Davies (2006) concluded, from their line strength
analysis, that secular evolution is not a dominant mechanism for
Sbc and earlier type spirals (see also Gonz\'alez-Delgado et al.\ 2004). 
MacArthur, Gonz\'alez \& Courteau (2009) revealed that most
bulges in all spiral galaxies have old mass-weighted ages, with
$<$25\% “by mass” of the stars being young.

Given that most bulges have old mass-weighted ages, it favours a prevalence of
classical bulges, many of which may co-exist with a secular-driven
pseudobulge, as advocated in Erwin et al.\ (2003) for the S0 population, 
and appears to also be the case for the Milky Way (D\'ek\'any et al.\ 2013).

\section{Compact massive bulges}\label{Sec6}

Fig.~\ref{Fig7} reveals that bulges are compact, with the smaller bulges being
similar to low-mass compact elliptical galaxies in the local universe and the
larger bulges being equivalent to the massive compact galaxies in the distant
universe (see Dullo \& Graham 2013).
Gas accretion from cold streams (Bouquin 2013; Combes 2013) 
is expected to build disks around the high-$z$ compact galaxies. 
This feeding is ultimately coplanar rather than random and thereby establishes
the disk (Pichon et al.\ 2011; Stewart et al.\ 2013; Prieto et al.\ 2013).  
As suggested in Graham (2013), and see Debattista et al.\ (2013) and 
Driver et al.\ (2013), the high-$z$ compact 
galaxies may indeed now be the old, massive bulges in today's 
large disk galaxies (Dullo 2013), while 
the local ($z=0$) compact elliptical galaxies may be the bulges of stripped disk 
galaxies, and/or some may be too small to have ever acquired a disk.

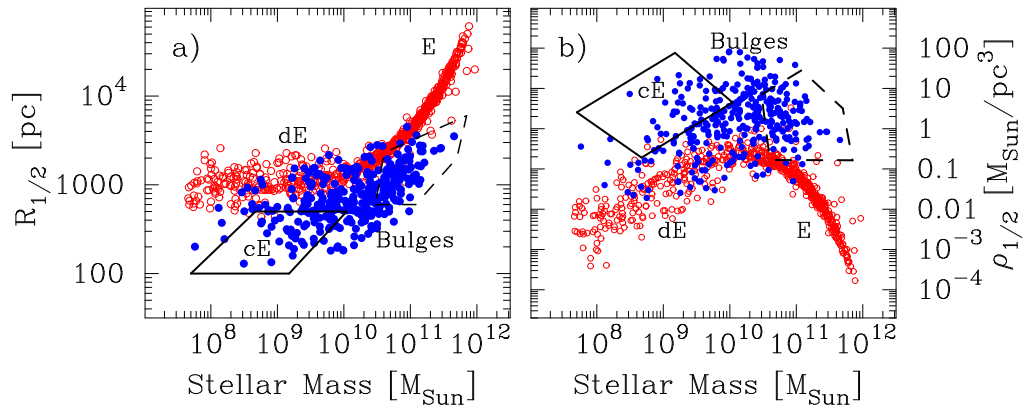
\articlefigure[angle=-90,scale=0.52]{grahama-fig7.ps}{Fig7}{Size-mass and density-mass diagrams,
  adapted from Graham (2013).  The single continuous, but curved, dwarf
  elliptical (dE) and elliptical (E) galaxy sequence is shown by the open
  circles.  The denser bulges are shown by the filled circles.  
  The dashed area shows the location of the
  compact, massive high-$z$ galaxies (Damjanov et al.\ 2009).  The solid
  rhomboidal shape denotes the location of compact elliptical (cE) galaxies
  observed in the local universe.}

\acknowledgements 
This research was supported under the Australian Research
Council’s funding scheme FT110100263.


\end{document}